\documentstyle[eqsecnum,aps]{revtex}
\begin{document}
\draft
\preprint{}
\def\be{\begin{equation}}
\def\ee{\end{equation}}
\def\bea{\begin{eqnarray}}
\def\eea{\end{eqnarray}}
\title{Quasi-long-range order in the random anisotropy Heisenberg
model: functional renormalization group in $4-\epsilon$ dimensions}
\author{D.E. Feldman}
\address{ Department of Condensed Matter Physics,
Weizmann Institute of Science, 76100 Rehovot, Israel\\
Landau Institute for Theoretical Physics,
 142432 Chernogolovka, Moscow region, Russia }
\maketitle

\begin{abstract}
The
large-distance behaviors of the random field and random
anisotropy $O(N)$ models are studied with the functional
renormalization group in $4-\epsilon$ dimensions. The random anisotropy
Heisenberg $(N=3)$ model is found to have a phase with an
infinite correlation length at low temperatures and weak disorder. The
correlation function of the magnetization obeys a power law $\langle
{\bf m}({\bf r}_1) {\bf m}({\bf r}_2)\rangle\sim| {\bf r}_1-{\bf
r}_2|^{-0.62\epsilon}$. The magnetic susceptibility diverges at low
fields as $\chi\sim H^{-1+0.15\epsilon}$.  In the random field $O(N)$
model the correlation length is found to be finite at the arbitrarily
weak disorder for any $N>3$.
The random field case is studied with a new simple method, based
on a rigorous inequality. This approach allows one
to avoid the integration
of the functional renormalization group equations.
\end{abstract}
\pacs{75.10 Nr, 75.50 Kj, 64.60 Cn}

\section{Introduction}

The effect of impurities on the order in condensed matter is
interesting, since the disorder is almost inevitably present in any
system. If the disorder is weak the short-range order is the same as
in the pure system. However, the large-distance behavior can be
strongly modified by the arbitrarily weak disorder. This happens in
the systems of continuous symmetry in the presence of the random
symmetry breaking field \cite{IM}. The first experimental example of
this kind is the amorphous magnet \cite{HPZ,OS}. During the last
decade a lot of other related objects were found. These are liquid
crystals in the porous media \cite{LQ}, nematic elastomers \cite{NE},
He-3 in aerogel \cite{He3} and vortex phases of
impure superconductors \cite{HTSC}. The
nature of the low-temperature phases of these systems is still
unclear. The only reliable statement is that a long-range
order is absent \cite{IM,L,P,AW}. However, other details
of the large-distance behavior are poorly understood.

The neutron scattering \cite{Xray} reveals sharp Bragg peaks in
impure superconductors at
low temperatures and weak external magnetic fields. Since the vortices
can not form a regular lattice \cite{L} it is tempting to assume
that there is a quasi-long-range order (QLRO), that is the correlation
length is infinite and correlation functions depend on the distance
slow. Recent theoretical \cite{RFXY,12a} and numerical \cite{numXY}
studies of the random field XY model, which is the simplest model of
the vortex system in the impure superconductor \cite{HTSC}, support
this picture. The theoretical advances \cite{RFXY,12a} are afforded by two
new technical approaches: the functional renormalization group
\cite{FRG} and the replica variational method \cite{MP}. These
methods are free from drawbacks of the standard renormalization group
and give reasonable results. The variational method regards the
possibility of spontaneous replica symmetry breaking and treats the
fluctuations approximately. On the other hand the functional
renormalization group provides a subtle analysis of the fluctuations
about the replica symmetrical ground state.  Surprisingly, the methods
give close and sometimes even the same results.

Both techniques were originally suggested for the random
manifolds \cite{FRG,MP} and then allowed to obtain information about
some other disordered systems with the Abelian symmetry
\cite{RFXY,12a,F,EN,RT}.
Less is known about the non-Abelian systems. The simplest of
them are the random field (RF) \cite{IM} and random anisotropy (RA)
\cite{HPZ} Heisenberg models. The latter was introduced as a model of
the amorphous magnet \cite{HPZ,OS}. In spite of a long discussion,
initiated by Ref. \cite{AP},
the question of QLRO in these models is still open. There is an
experimental evidence in favor of no QLRO  \cite{B}.
On the other hand recent numerical simulations \cite{num} support the
possibility of QLRO in these systems. The only theoretical approach,
developed up to now, is based on the spherical approximation
\cite{P,spher,G}
and predicts the absence of QLRO at $N\gg 1$ magnetization components.
However, there is no reason for this approximation to
be valid at $N\sim 1$.

In this paper we study the RF and RA
$O(N)$ models in $4-\epsilon$ dimensions with the functional
renormalization group. The large-distance behaviors of the systems are
found to be quite different.  Whereas in the RF $O(N)$ model
with $N>3$ the correlation length is always finite, the RA
Heisenberg $(N=3)$ model has a phase with QLRO. In that phase the
correlation function of the magnetization obeys a power law and the
magnetic susceptibility diverges at low fields.

The paper has the following structure. In the second section
the models are formulated and a qualitative discussion is given.
The third section contains a derivation
of the one-loop renormalization group
(RG) equations. The 4th section is devoted to the RF model.
The absence of QLRO in that model at $N>3$ is shown with
a new method, based on a rigorous inequality. This approach
simplifies tedious RG calculations
and can be useful in other problems.
The RA case is considered in the 5th section.
The stable RG fixed point corresponds to QLRO in $4-\epsilon$
dimensions at $N\sim 1$.
In particular, at the weak disorder the correlation length is infinite
in the low temperature phases of the RA XY ($N=2$)
and Heisenberg ($N=3$) models.  However, QLRO is absent at $N\ge 10$.
In 4 dimensions the correlation functions of the RA Heisenberg model
depend on the distance logarithmically. The exact result for the
two-spin correlator is given by the $\ln^{-0.62}R$ law.
The Conclusion contains a
discussion of the results.  Appendix A is devoted to a generalization
of the Schwartz-Soffer inequality \cite{SwSo}. The generalized
inequality is applied to the stability analyses of the RG fixed
points.  Appendix B describes a simple Migdal-Kadanoff renormalization
group approach that reproduces qualitatively the results of the rigorous
method. This approximation provides good estimations of the
critical exponents of the RA XY and Heisenberg models.
Appendix C includes some technical details of the functional
RG in the spherical model.

\section{Model}

To describe the large-distance behavior at low temperatures we use the
classical nonlinear $\sigma$-model with the Hamiltonian

\be
\label{1}
H=\int d^D x[J\sum_\mu\partial_\mu{\bf n}({\bf x})
\partial_\mu{\bf n}({\bf x}) + V_{\rm imp}({\bf x})],
\ee
where ${\bf n}({\bf x})$ is the unit vector of the magnetization,
$V_{\rm imp}({\bf x})$ the random potential. In the RF case
it has the form

\be
\label{2}
V_{\rm imp}=-\sum_\alpha h_\alpha({\bf x})n_\alpha({\bf x});
\alpha=1,...,N,
\ee
where the random field ${\bf h}({\bf x})$ has a Gaussian
distribution and $\langle h_\alpha({\bf x})h_\beta({\bf
x}')\rangle=A^2\delta({\bf x}-{\bf x}')\delta_{\alpha\beta}$. In the
RA case the random potential is given by the equation

\be
\label{3}
V_{\rm imp}=-\sum_{\alpha,\beta}\tau_{\alpha \beta}({\bf
x})n_\alpha({\bf x})n_\beta({\bf x}); \alpha,\beta=1,...,N,
\ee
where $\tau_{\alpha\beta}({\bf x})$ is a Gaussian random variable,
$\langle\tau_{\alpha\beta}({\bf x})\tau_{\gamma\delta}({\bf
x}')\rangle=A^2\delta_{\alpha\gamma}\delta_{\beta\delta}\delta({\bf
x}-{\bf x}')$. The random potential (\ref{3}) corresponds to the
same symmetry as the more conventional choice
$V_{\rm imp}=-({\bf hn})^2$ but is more convenient for the further
discussion.

We assume that the temperature is low and
the thermal fluctuations are negligible.
The Imry-Ma argument \cite{IM,P} suggests that in our problem the
long-range order is absent at any dimension $D<4$.
One can estimate the
Larkin length, up to which there are strong ferromagnetic
correlations, with the following qualitative
RG approach.
Let one remove the fast modes and rewrite the
Hamiltonian in terms of the block spins, corresponding to the scale
$L=ba$, where $a$ is the ultraviolet cut-off, $b>1$. 
Then let one make
rescaling such that the Hamiltonian would restore its initial form with
new constants $A(L), J(L)$. Dimensional analysis provides estimations

\be
\label{4}
J(L)\sim b^{D-2} J(a); A(L)\sim b^{D/2}A(a)
\ee
To estimate the typical angle $\phi$ between neighbor block spins, one
notes that the effective field, acting on each spin, has two
contributions: the exchange contribution and the random one. The
exchange contribution of order $J(L)$ is oriented along the local
average direction of the magnetization. The random contribution of
order $A(L)$ may have any direction. This allows one to write at low
temperatures that $\phi(L)\sim A(L)/J(L)$. The Larkin length
corresponds to the condition $\phi(L)\sim 1$ and equals $L\sim
(J/A)^{2/(4-D)}$ in agreement with the Imry-Ma argument \cite{IM}.
If Eq. (\ref{4}) were exact the Larkin length could be interpreted as
the correlation length. However, there are two sources of
corrections to Eq. (\ref{4}). Both of them are relevant already at the
derivation of the RG equation for the pure system in $2+\epsilon$
dimensions \cite{Pol}. The first source is the renormalization due to
the interaction and the second one results from the
magnetization rescaling
which is necessary to ensure the fixed length condition
${\bf n}^2=1$. The leading corrections to Eq. (\ref{4}) are
proportional to $\phi^2 J, \phi^2 A$. Thus, the RG equation for the
combination $(A(L)/J(L))^2$ reads

\be
\label{6}
\frac{d}{d \ln L}\left(\frac{A(L)}{J(L)}\right)^2=
\epsilon\left(\frac{A(L)}{J(L)}\right)^2+
c\left(\frac{A(L)}{J(L)}\right)^4, \epsilon=4-D
\ee
If the constant $c$ in Eq. (\ref{6}) is positive the Larkin length is
the correlation length indeed. But if $c<0$ the RG equation has a fixed
point, corresponding to the phase with an infinite correlation
length. As seen below, both situations are possible, depending
on the system.

The large-distance behaviors of the RF and RA $O(N)$ models
are known in
two limit cases: $N=2$ and $N=\infty$. In the spherical limit
($N=\infty$) QLRO is absent (Appendix C, \cite{spher}) while
the XY model possesses QLRO \cite{RFXY,12a,prim}.  Hence,
the ordering disappears at some critical number $N_c$ of the
magnetization components. This critical number is larger in the RA
model, since the fluctuations of the magnetization are stronger in the
RF case. Indeed, in the RF model the magnetization tends to be oriented
along the random field, whereas in the RA case there are two preferable
local magnetization directions so that the spins tend to lie in the
same semispace.

\section{RG equations}

In the previous section the RG equations are discussed from
the qualitative point of view. Eq. (\ref{6}) corresponds
to the Migdal-Kadanoff approach of Appendix B. In the present section
we derive the RG equations in a systematic way.

The one-loop RG equations for the $N$-component RF and RA models in
$4+\epsilon$ dimensions were already derived in Ref. \cite{DF}. We can
directly use that result, since the RG equations in dimensions $D<4$ can
be obtained by just changing the sign of $\epsilon$. However, the
approach \cite{DF} is cumbersome and we provide below a simpler
derivation.

We use the method, suggested by Polyakov \cite{Pol} for the pure system
in $2+\epsilon$ dimensions. This method is technically simpler and
closer to the intuition than the other approaches.
A disadvantage of the method is the difficulty of the
generalization for the higher orders in $\epsilon$.
This generalization requires the field-theoretical approach
\cite{ZJ}.

The same consideration as in the XY
\cite{12a} and random manifold \cite{FRG} models suggests that
near a
zero-temperature fixed point in $4-\epsilon$ dimensions there is an
infinite set of relevant operators.
Let us show that
after the replica averaging the relevant part of the effective replica
Hamiltonian can be represented in the form

\be
\label{7}
H_R=\int d^D x[\sum_a\frac{1}{2T}\sum_\mu
\partial_\mu{\bf n}_a\partial_\mu{\bf n}_a - \sum_{ab}\frac{R({\bf
n}_a{\bf n}_b)}{T^2}],
\ee
where $a,b$ are replica indices, $R(z)$ is some function, $T$ the
temperature.
We ascribe to the field ${\bf n}$ the scaling dimension $0$.
We also assume that the coefficient before the
gradient term in (\ref{7}) is $1/(2T)$ at any scale.
Then in the $(4-\epsilon)$-dimensional space
the scaling dimension of the temperature
$\Delta_T=-2+O(\epsilon)$. Any operator $A$ containing
$m$ different replica indices is proportional \cite{FRG}
to $1/T^m$. Hence,
the scaling dimension $\Delta_A$ of the operator $A$
satisfies the relation $\Delta_A=4-n+m\Delta_T+O(\epsilon)$, where $n$
is the number of the spatial derivatives in the operator.
The relevant operators have $\Delta_A\ge 0$.
Hence, the relevant operators
can not contain more than two different replica
indices. A symmetry consideration shows that all the possible relevant
operators are included into Eq. (\ref{7}).
The function $R(z)$ is arbitrary in the RF case.
In the RA case $R(z)$ is even
due to the symmetry with respect to changing the sign of the
magnetization.

The one-loop RG equations for the $N$-component model
in $4-\epsilon$ dimensions are obtained by
a straightforward combination of the methods of Refs. \cite{FRG} and
\cite{Pol}. We express each replica
${\bf n}^a({\bf x})$
of the magnetization
as a combination of
fast
fields $\phi_i^a({\bf x}), i=1,...,N-1$
and a slow field ${\bf n}'^a({\bf x})$
of the unit length.
We use the representation

\be
\label{dec}
{\bf n}^a({\bf x})={\bf n}'^a({\bf x})\sqrt{1-\sum_i(\phi_i^a({\bf
x}))^2}+ \sum_i\phi_i^a({\bf x}){\bf e}_i^a({\bf x}),
\ee
where the unit vectors ${\bf e}_i^a({\bf x})$ are perpendicular
to each other and the vector ${\bf n}'^a({\bf x})$.
The slow field ${\bf n}'^a$ can be chosen in different ways.
For example, one can cut the system into blocks of size
$L\gg a$, where $a$ is the ultra-violet cut-off.
In the center ${\bf x}$ of a block the vector
${\bf n}'^a({\bf x})$ should be
parallel to the total magnetization of the block.
In the other points the field ${\bf n}'^a$ should
be interpolated.
We assume that the fluctuations of the magnetization are
weak, that is $\langle\phi_i^2\rangle\ll 1$.
Then the fluctuations of the field ${\bf n}^a$ are orthogonal to the
vector ${\bf n}'^a$
because of the fixed length constraint $({\bf n}^a)^2=1$.

To substitute the expression
(\ref{dec}) into the Hamiltonian we have to differentiate the vectors
${\bf e}_i^a$. We use the following definition

\be
\label{dife}
\frac{\partial {\bf e}^a_i}{\partial x_{\mu}}=
-c^a_{\mu i}{\bf n}'^a+\sum_k f^a_{\mu, ik}{\bf e}_k^a.
\ee
It is easy to show \cite{Pol} that $\sum_{\mu i}(c^a_{\mu i})^2=
\sum_{\mu}(\partial_{\mu}{\bf n}'^a)^2$. With the accuracy up to the
second order in $\phi$ the replica Hamiltonian (\ref{7}) can be
represented as follows

\bea
\label{Hphi}
H_{R}=\int d^D x
[
\frac{1}{2T}\sum_{a}
\{
(\partial_{\mu}{\bf n}'^a)^2
(1-(\phi_i^a)^2)+
c^a_{\mu i}c^a_{\mu k}\phi^a_i\phi^a_k  +
( \partial_{\mu}
\phi_i^a -
f^a_{\mu, ik}
\phi^a_k)^2
\}
& & \nonumber \\
- \frac{1}{T^2}\sum_{ab}
\{
R({\bf n}'^a{\bf n}'^b)+A^{ab}(\phi^a_i)^2+
B^{ab}_{ij}\phi^a_i\phi^a_j +
C^{ab}_{ij}\phi_i^a\phi^b_j
\}
], & &
\eea
where the summation over the repeated indices $i,j,k,\mu$ is assumed
and

\bea
\label{coef}
A^{ab}=-({\bf n}'^a{\bf n}'^b)R'({\bf n}'^a{\bf n}'^b);
B^{ab}_{ij}=
({\bf n}'^b{\bf e}^a_i)({\bf n}'^b{\bf e}^a_j)
R''({\bf n}'^a{\bf n}'^b); & & \nonumber \\
C^{ab}_{ij}=
({\bf e}^a_i{\bf e}_j^b)R'({\bf n}'^a{\bf n}'^b)
+
({\bf n}'^a{\bf e}^b_j)({\bf n}'^b{\bf e}^a_i)
R''({\bf n}'^a{\bf n}'^b). & &
\eea
In the last formula $R'$ and $R''$ denote the first and second
derivatives of the function $R(z)$.
We have omitted the terms of the first order in $\phi$
in Eq. (\ref{Hphi}). These terms are proportional to the products of the
fast field $\phi$ and some slow fields. Hence, they are
non-zero only in narrow shells of the momentum space.
One can show that their contributions to the RG equations are
negligible.

To obtain the RG equations we have to integrate out the fast
variables $\phi^a_i$.
Near a zero-temperature fixed point the Jacobian of the transformation
${\bf n}\rightarrow ({\bf n}', \phi_i)$ can be ignored,
since the Jacobian does not depend on the temperature.
The integration measure is determined from the condition that
the field ${\bf n}'^a$ is a slow part of the magnetization.
This condition imposes restrictions on the fields $\phi$.
The expression (\ref{Hphi}) depends
on the choice of the vectors ${\bf e}^a_i$ (\ref{dec}). However,
after integrating out the fields $\phi$ the Hamiltonian can depend only
on the slow part ${\bf n}'^a$
of the magnetization.
One can
make the calculations simpler, considering special realizations of the
field ${\bf n}'^a$. To find the renormalization of the disorder-induced
term $R(z)$ (\ref{7}) we can assume that the field ${\bf n}'^a$ does not
depend on the coordinates. The renormalization of the gradient energy
can be determined, assuming that the vectors ${\bf n}'^a({\bf x})$
depend on one spatial coordinate only
and lie in the same plane. In both cases the
vectors ${\bf e}_i^a$ can be chosen such that in Eq. (\ref{dife})
$f^a_{\mu, ik}=0$. At such a choice the
integration measure can be omitted
and the fields $\phi_i^a$ can be considered as
weakly interacting fields
with the wave vectors from the interval $1/a>q>1/L$.

To derive the one-loop RG equations we express
the free energy via the Hamiltonian (\ref{Hphi}).
Then we expand the exponent in the partition function up to
the second order in
$(H_R-\int d^D x \sum_{\mu i}(\partial_{\mu}\phi_i)^2/(2T))$
and integrate over $\phi^a_i$. Finally, we make a rescaling.
The vectors ${\bf e}^a_i$
can be excluded from the final expressions with the relation
$\sum_i({\bf ae}^a_i)({\bf be}^a_i)=
({\bf ab})-({\bf an}'^a)({\bf bn}'^a)$.
In a zero-temperature fixed point the RG equations are

\be
\label{Tz}
\frac{d\ln T}{d\ln L}= -(D-2) + 2(N-2)R'(1)+O(R^2,T);
\ee
\begin{eqnarray}
\label{Rz}
0=\frac{dR(z)}{d \ln L}=\epsilon R(z) + 4(N-2)R(z)R'(1)-
2(N-1)zR'(1)R'(z)
+2(1-z^2)R'(1)R''(z) & & \nonumber \\
+
(R'(z))^2(N-2+z^2)-2R'(z)R''(z)z(1-z^2)+
(R''(z))^2(1-z^2)^2,  & &
\end{eqnarray}
where the factor $1/(8\pi^2)$ is absorbed into $R(z)$ to simplify
notations.
The RG equations become simpler after the substitution of the
argument of the function $R(z)$: $z=\cos\phi$. In terms of this new
variable one has to find even periodic solutions $R(\phi)$. The period
is $2\pi$ in the RF case and $\pi$ in the RA case due to the additional
symmetry of the RA model.
The one-loop equations get the form

\be
\label{8}
\frac{d\ln T}{d\ln L}= -(D-2) - 2(N-2)R''(0)+O(R^2,T);
\ee
\begin{eqnarray}
0=\frac{dR(\phi)}{d \ln L}=\epsilon R(\phi) + (R''(\phi))^2 - 2R''(\phi)
R''(0) -  & & \nonumber\\
\label{9}
(N-2)[4R(\phi)R''(0)+2{\rm ctg}\phi R'(\phi) R''(0) -
\left(\frac{R'(\phi)}{\sin\phi}\right)^2] + O(R^3,T) & &
\end{eqnarray}
Eq. (\ref{8}) provides the following result for the
scaling dimension $\Delta_T$ of the temperature

\be
\label{DT}
\Delta_T=-2+\epsilon-2(N-2)R''(0).
\ee

The two-spin correlation function is given in the one-loop
order \cite{Pol} by the expression

\be
\label{cf}
\langle{\bf n}^a({\bf x}){\bf n}^a({\bf x}')\rangle=
\langle{\bf n}'^a({\bf x}){\bf n}'^a({\bf x}')\rangle
(1-\langle\sum_i(\phi_i^a)^2\rangle).
\ee
Hence, in the fixed point
$\langle{\bf n}({\bf x}){\bf n}({\bf x}')\rangle\sim|{\bf x}-{\bf
x}'|^{-\eta}$, where

\be
\label{11}
\eta=-2(N-1)R''(\phi=0)
\ee

Let us find the magnetic susceptibility
in the weak uniform external field $H$. We add to the Hamiltonian
(\ref{7})
the term $-\sum_a\int d^Dx Hn^a_z/T$
(the field is directed along the z-axis).
The renormalization of the field $H$ is determined by the
renormalization of the temperature (\ref{8}) and the field ${\bf n}$.
In the zero-loop order
the renormalized magnetic field $h(L)$
depends on the scale as $h(L)=H\times(L/a)^2$.
Hence, the correlation length $R_c\sim H^{-1/2}$.
The magnetization, averaged over a block of size $R_c$,
is oriented along the field. The absolute value of this average
magnetization is proportional to $R_c^{-\eta/2}$.
This allows us to calculate
the critical exponent $\gamma$ of the magnetic susceptibility
$\chi(H)\sim H^{-\gamma}$ in a zero-temperature fixed point:

\be
\label{15}
\gamma=1+(N-1)R''(\phi=0)/2 .
\ee

In Ref. \cite{DF}
Eqs. (\ref{Tz},\ref{Rz}) were derived with a different
method. In that paper the critical behavior in
$4+\epsilon$ dimensions was studied by considering analytical
fixed point solutions $R(z)$. In the Heisenberg model, analytical
solutions are absent and they are unphysical for $N\ne 3$ \cite{DF}.
In $4-\epsilon$ dimensions appropriate analytical solutions are absent
for any $N$.  To demonstrate this
let us differentiate Eq. (\ref{Rz}) over $z$ at $z=1$.  For any
analytical $R(z)$ we obtain the following flow equation

\be
\label{flow}
\frac{dR'(z=1)}{d \ln L}=\epsilon R'(z=1) + 2(N-2)(R'(z=1))^2.
\ee
At $N> 2$ the fixed point of this equation
$R'(z=1)=-\epsilon/[2(N-2)]<0$. It corresponds to the
negative critical exponent $\eta$ (\ref{11}) and hence is
unphysical. However, we shall see that in the RA model
some appropriate non-analytical fixed points $R(z)$ appear. In these
fixed points $R''(z=1)=\infty$.  In Ref. \cite{DF} the RG charges are
the derivatives of the function $R(z)$ at $z=1$. Thus, in a
non-analytical fixed point these charges diverge. In the systems with a
finite number of the charges their divergence implies the absence of a
fixed point.  However, if the number of
the RG charges is infinite such a  criterion does not work
 and is even ambiguous. Indeed, the set
of charges can be chosen in different ways and
e.g. the coefficients of the Taylor expansion about $z=0$
remain finite in our problem.

\section{Random field}
\label{sec:IV}

For the RF XY model the one-loop RG equations (\ref{8},\ref{9})
can be solved exactly
\cite{12a}. The solution
corresponds to QLRO with the critical exponents
$\eta=\pi^2/9\epsilon, \gamma=1-\pi^2/18\epsilon$.
In the first order in $\epsilon$ the exponent $\eta$
equals the prefactor $C$ before the logarithm
in the correlation function \cite{12a}
of the angles $\phi({\bf x})$ between the spins
${\bf n}({\bf x})$ and some fixed direction:
$\langle(\phi({\bf x}_1)-\phi({\bf x}_2))^2\rangle=C\ln|{\bf x}_1-{\bf
x}_2|$.
We expect that this coincidence
does not extend to the higher orders.

If $N\ne 2$ the RG equation (\ref{9}) is more complex.
Fortunately, at $N>3$ there is still a simple method
to study the large-distance behavior. The method is
based on the Schwartz-Soffer inequality \cite{SwSo} and shows that
QLRO is absent.

In Ref. \cite{SwSo} the inequality is proven for the
Gaussian distribution of the random field.
It can also be proved for the arbitrary RF distribution
(Appendix A).

Let us demonstrate the absence of physically acceptable fixed points
in the RF case at $N>3$. We derive some inequality for
critical exponents. Then we show that the inequality has no solutions.
We use a rigorous inequality for the connected and
disconnected correlation functions \cite{SwSo}

\be
\label{17}
\langle\langle{\bf n}({\bf q}){\bf n}(-{\bf q})
\rangle\rangle =
\langle{\bf n}_a({\bf q}){\bf n}_a(-{\bf q})\rangle -
\langle{\bf n}_a({\bf q}){\bf n}_b(-{\bf q})\rangle \le
{\rm const}\sqrt{\langle{\bf n}_a({\bf q}){\bf n}_a(-{\bf q})\rangle},
\ee
where ${\bf n}({\bf q})$ is a Fourier-component of the magnetization,
$a,b$ are replica indices.
The disconnected correlation function is described by
the critical exponent (\ref{11}).
The large-distance
behavior of the connected correlation function in a zero-temperature
fixed point can be derived from the expression

\be
\label{eqchi}
\chi\sim\int\langle\langle {\bf n}({\bf 0}){\bf n}({\bf
x})\rangle\rangle d^D x
\ee
and the critical exponent of the susceptibility (\ref{15}).
The integral in the right hand side of Eq. (\ref{eqchi})
is proportional to $R_c^{D-\eta_1}$, where
$R_c$ is the correlation length in the external field $H$,
$\eta_1$ the critical exponent of the connected correlation
function. For the calculation of the exponent $\gamma$
(\ref{15}) we used the
zero-loop expression of $R_c$ via $H$. Now we need the one-loop
accuracy. In this order $R_c\sim H^{-1/[2-(N-3)R''(0)]}$.
This allows us to get the following equation for the exponent $\eta_1$

\be
\label{e1}
\eta_1=D-2-2R''(0).
\ee
In a fixed point Eq. (\ref{17}) provides
an inequality for the critical exponents of the connected and
disconnected correlation functions \cite{SwSo}.
The inequality has the form

\be
\label{ein}
2(2-D+\eta_1)\ge 4-D+\eta.
\ee
This allow us to obtain the following relation

\be
\label{18}
4-D \le\frac{3-N}{N-1}\eta+o(R),
\ee
where $\eta$ is given by Eq. (\ref{11}).
The two-spin correlation function can not increase
up to the infinity as the distance
increases. Hence, the critical exponent $\eta$ is positive.
At $N>3$ this is incompatible with Eq. (\ref{18})
at small $\epsilon$.
Thus, there are no accessible fixed points for
$N>3$. This suggests the strong coupling regime with a
presumably finite correlation length.

Certainly,
in the RF XY model \cite{RFXY,12a} Eq. (\ref{18}) is satisfied.
However, the unstable fixed points of the RG equations
\cite{12a} do not satisfy the inequality.

The marginal Heisenberg case $N=3$ is the most difficult,
since in the one-loop order the right hand side of
Eq. (\ref{18}) equals zero at $N=3$. Hence, the two-loop
corrections may be relevant.
The RF Heisenberg model is beyond
the scope of the present paper.

\section{Random anisotropy}

In this section we investigate the possibility of QLRO in the RA $O(N)$
model. The first subsection is devoted to the simplest case of
the XY model. The second subsection contains an inequality for
the critical exponent $\eta$. The derivation of the inequality
is analogous to Eq. (\ref{18}). This inequality is applied
in the next subsections. The third subsection contains
the results for the Heisenberg model.
In the last subsection we consider the case
$N>3$.

\subsection{$N=2$}

This case is studied analogously to the RF XY model \cite{12a}.
At $N=2$ the RG equation (\ref{9}) can be solved analytically.
Its solution is a periodical function with period $\pi$. In
interval $0<\phi<\pi$ the fixed point solution
$R(\phi)$ is given by the formula

\be
\label{RAXYsol}
R(\phi)=\frac{\pi^4\epsilon}{144}\left[1/36-({\phi}/{\pi})^2
\left(1-({\phi}/{\pi})\right)^2\right].
\ee
It is a stable fixed point. This can be
verified with the linearization of the flow equation (\ref{9})
for the small
deviations from the fixed point.
Another proof of the stability is
based on the inequality of the next subsection.

The stable fixed point corresponds to the QLRO phase at low temperatures
and weak disorder. The critical exponents $\eta=\pi^2\epsilon/36,
\gamma=1-\pi^2\epsilon/72$.

The solution (\ref{RAXYsol}) is non-analytical at $\phi=0$,
since $R^{IV}(\phi=0)=\infty$.
Hence, the Taylor expansion over $\phi$ is absent.
However, a power expansion over $|\phi|$ exists. We shall see below
that the same behavior at small $\phi$ conserves also at other
$N$.

\subsection{An inequality for a critical exponent}
\label{sec:V.B}

We use the same approach as in the RF model.
Since in the RA case the random field is conjugated
with a second order polynomial of the magnetization,
the Schwartz-Soffer inequality \cite{SwSo} should be
applied to correlation functions of
the field $m({\bf x})=(n_z({\bf x}))^2-1/N$, where $n_z$
denotes one of the magnetization components, $1/N$ is subtracted
to ensure the relation $\langle m\rangle=0$.

To calculate the critical exponent $\mu$ of the disconnected correlation
function we use the representation (\ref{dec})
and obtain the relation

\be
\label{RGm}
\langle m^a({\bf x})m^a({\bf x}')\rangle=
\langle m'^a({\bf x})m'^a({\bf x}')\rangle\left(1-\frac{2N\sum_i
\langle(\phi_i^a)^2\rangle}{N-1}\right),
\ee
where $a$ is a replica index, $m'=(n'_z)^2-1/N$
the slow part of the field $m$. One finds $\mu=-4NR''(0)$.

The critical exponent $\mu_1$ of the connected correlation function
is determined analogously to the RF case. We apply a weak
uniform field $\tilde H$, conjugated with the field $m$, and
calculate the susceptibility $dm/d\tilde H$ in two ways.
The result for the critical exponent is $\mu_1=D-2-2(N+2)R''(0)$.

The Schwartz-Soffer inequality provides a relation between
the exponents $\mu$ and $\mu_1$.
It has the same structure as Eq. (\ref{ein}).
Finally, we obtain the following equation

\be
\label{RAineq}
\eta\ge\frac{4-D}{4}(N-1)+o(R).
\ee
In terms of the RG charge $R(\phi)$ this inequality can be rewritten
in the form
\be
\label{Rin}
R''(0)\le-\epsilon/8+o(R).
\ee

\subsection{$N=3$}
\label{sec:V.C}

In this case we solve Eq. (\ref{9}) numerically.
Since coefficients of Eq. (\ref{9}) are
large as $\phi\rightarrow 0$, it is convenient to use a series expansion
of the fixed-point solution
$R(\phi)$ at small $\phi$.  At the larger $\phi$ the
equation can be integrated with the Runge-Kutta method.
The following expansion over $t=\sqrt{(1-z)/2}=|\sin(\phi/2)|$ holds

\bea
\label{smphi}
R(\phi)/\epsilon
=\frac{(N-1)a^2}{1-4(N-2)a}+2a\sin^2\frac{\phi}{2}
\pm\frac{4\sqrt{2}}{3}\sqrt{\frac{-a+2(N-2)a^2}{N+2}}|\sin^3\frac{\phi}{2}|
& & \nonumber\\
+\left(\frac{2a}{3}-\frac{2}{3(N+4)}\right)\sin^4\frac{\phi}{2}+
O(|\sin^5\frac{\phi}{2}|), & &
\eea
where $a=R''(\phi=0)/\epsilon$. We see that the RG charge $R(\phi)$ is
non-analytical at small $\phi$. Similar to the random manifold
\cite{FRG} and random field XY \cite{12a} models $R^{IV}(0)=
\infty$.

Numerical calculations show that at any $N$ the solutions,
compatible with the inequality (\ref{Rin}),
have sign "$+$" before the third term of Eq. (\ref{smphi}).
The solutions to be found are even periodical functions with period $\pi$.
Hence, their derivative is zero at $\phi=\pi/2$.
At $N=3$ there is only one solution that satisfies Eq. (\ref{Rin}).
It corresponds to $R''(\phi=0)=-0.1543\epsilon$.
If this solution is stable Eqs. (\ref{11},\ref{15}) provide
the following results for the critical exponents

\be
\label{crexp}
\eta=0.62\epsilon; \gamma=1-0.15\epsilon.
\ee
All the other solutions of Eq. (\ref{9})
do not satisfy Eq. (\ref{Rin}) and hence are unstable.

We have still to test the stability of the solution found.
For this aim we use an approximate method. First, we find an
approximate analytical solution of Eq. (\ref{9}).
We rewrite Eq. (\ref{9}), substituting $\omega(R''(\phi))^2$ for
$(R''(\phi))^2$.  The case of interest is $\omega=1$ but at $\omega=0$
the equation can be solved exactly. The solution at $\omega=1$ can
then be found with the perturbation theory over $\omega$. The exact
solution at $\omega=0$ is $R_{\omega=0}(\phi)=\epsilon(\cos
2\phi/24+1/120)$. The corrections of order $\omega^k$ are
trigonometric polynomials of order $2(k+1)$. The first correction
is

\be
\label{fc}
R_1(\phi)=-\frac{2\omega\epsilon}{99}\cos 2\phi
+\frac{\omega\epsilon}{264}\cos 4\phi
+{\rm const}
\ee
After the calculation of the corrections we can write
an asymptotic series for the critical exponent $\eta$ (\ref{11}):
$\eta=\epsilon(0.67-0.08\omega+0.14\omega^2-\dots)$. The resulting
estimation $\eta=\epsilon(0.67\pm0.08)$ agrees with the numerical result
(\ref{crexp}) well.
This allows us to expect that the stability analysis
of the solution $R_{\omega=0}$ of the equation with $\omega=0$ provides
information about the stability of the solution of Eq. (\ref{9}).

To study the stability of the exact solution of the equation
with $\omega=0$ is a simple problem. We introduce a small deviation
$r(\phi)$: $R(\phi)=R_{\omega=0}(\phi)+r(\phi)$ and write the
flow equation for this deviation:

\be
\label{fl}
\frac{dr(\phi)}{d\ln L}=(5r(\phi)+r''(\phi)+
r''(0)\cos2\phi)/3+{\rm const}\times r''(0).
\ee
It is convenient to use the Fourier expansion $r(\phi)=
\sum_m a_m\cos2m\phi$. The flow equations for the Fourier
harmonics can be easily integrated. We see that $a_m\rightarrow 0$
as $L\rightarrow\infty$ for any $m>0$. The solution is unstable
with respect to the constant shift $a_0$. However, this instability
has no interest for us, since the correlation functions
do not change at such shifts \cite{FRG}.
Indeed, the constant shift
corresponds to the addition of just a random term, independent of
the magnetization, to the Hamiltonian (\ref{1}).
Thus, the RG equation possesses a stable fixed point.
This fixed point describes the QLRO phase with the critical exponents
(\ref{crexp}).

In the Abelian systems the results of the functional RG are supported
by the variational method \cite{MP}. In our problem this method can not
be applied. However, it is interesting that in the Abelian systems
the functional RG equations without $(R''(\phi))^2$ reproduce
the variational results.

As usual in critical phenomena, in $4$ dimensions the one-loop RG
equations allow one to obtain the exact large-distance asymptotics of
the correlation function. In the 4-dimensional case
$R(\phi)=\tilde R(\phi)/\ln L$, where
$\tilde R(\phi)$ satisfies Eq. (\ref{9}) at $\epsilon=1$.
We obtain the following result for the two-spin
correlation function with Eq. (\ref{cf})

\be
\label{elg}
\langle{\bf n}({\bf x}){\bf n}({\bf x}')\rangle\sim\ln^{-0.62}|{\bf
x}-{\bf x}'|.
\ee

\subsection{$N>3$}

Numerical analysis of Eq. (\ref{9}) shows that solutions,
compatible with Eq. (\ref{Rin}), are absent at $N\ge 10$.
Hence, QLRO is absent for any $N\ge 10$.
In the spherical model ($N=\infty$) the absence of fixed points can be
demonstrated analytically (Appendix C). This agrees with
the previous results \cite{P,spher}.
For each integer $N<10$ the RG equation (\ref{9})
has exactly one solution, satisfying the inequality
of section \ref{sec:V.B}. These solutions are described
in Table I. In the table, $\eta$ is the critical exponent
of the two-spin correlation function, $\Delta_T$ the
scaling dimension of the temperature (\ref{DT}).

Unfortunately, it is not clear if the fixed points, found at $N>3$,
survive in 3 dimensions. A zero-temperature
fixed point can exist only if the scaling dimension of the
temperature is negative. Table \ref{table1} shows that scaling
dimension is positive in the one-loop approximation at $\epsilon=1$ and
$N\ge 5$. In the 3-dimensional $O(4)$ model the one-loop correction to
the scaling dimension $-2(N-2)R''(0)\approx 0.7\epsilon$ is close to the
zero-loop approximation $-2+\epsilon$.  Thus, the next orders of the
perturbation theory are crucial to understand what happens in 3
dimensions.

In the $O(2)$ model the scaling dimension $\Delta_T=-2+\epsilon$ is
exact \cite{12a,FRG}. Hence, QLRO disappears in 2 dimensions.
In  systems with a larger numbers of magnetization
components fluctuations become stronger. Thus, one expects
the absence of QLRO in all the two-dimensional $O(N)$ models.

At the zero temperature Eq. (\ref{9}) is valid
independently of the scaling dimension
$\Delta_T$.
It is tempting to assume that at the zero temperature
QLRO still exists in the RA $O(N>3)$ models below the
critical dimension, in which $\Delta_T=0$.
However,
the experience of the two-dimensional RF XY model
does not support such an expectation.
Recent numerical simulations show that QLRO is absent
even in the ground state of that model \cite{2DGSsim}.

\begin{table}
\caption{Critical exponents of the RA $O(N)$ model.}
\label{table1}
\begin{tabular}{ccccccccc}
$N$ & 2 & 3 & 4 & 5 & 6 & 7 & 8 & 9 \\
\tableline
$\eta$ & $\pi^2\epsilon/36$ &
$0.62\epsilon$ & $1.1\epsilon$ & $1.7\epsilon$ & $2.7\epsilon$ &
$4.6\epsilon$ & $9.0\epsilon$ & $33\epsilon$ \\
$\Delta_T$ & -2+$\epsilon$ & -2+$1.3\epsilon$ & -2+$1.7\epsilon$ &
-2+$2.3\epsilon$ & -2+$3.2\epsilon$ & -2+$4.8\epsilon$ &
-2+$8.7\epsilon$ & -2+$30\epsilon$ \\
\end{tabular}
\end{table}

\section{Conclusion}

We have obtained QLRO in the RA Heisenberg model.
This is the first example of QLRO in a non-Abelian system.
The RF disorder tends to destroy the ordering which exists in the RA
case. This difference between the RF and RA models is not surprising,
since the same difference was already obtained in Ref. \cite{F} for
the two-dimensional RF and RA XY models with the dipole forces.

We have not yet discussed the role of the topological defects.
The contribution of the topological excitations to the RG equations
(\ref{8},\ref{9}) is determined by the rare regions where the
random field is sufficiently strong to compensate the core energy.
Hence,
similar to the pure system in $2+\epsilon$ dimensions they are
responsible for the non-perturbative corrections of order
$\exp(-1/\epsilon)$. Thus, their effect is negligible
at small $\epsilon$. Several studies were devoted to
the role of the vortices in the RF XY model \cite{vortex}.
The theoretical prediction of QLRO in this system
is based on the vortexless version of the model \cite{RFXY,12a}.
A qualitative estimation \cite{12a} and variational calculations
\cite{vortex} suggest that the topological defects do not
change the behavior of the RF XY model at the weak disorder.
Our approach allows us to consider the XY model, including vortices. We
see that QLRO does exist in the model with the defects.

However, in our problem there may be a more important source
of the non-perturbative corrections. The
effect of the multiple energy minima can lead to corrections
of order $\epsilon^{5/2}$ to the RG equations \cite{FRG}.
Unfortunately, the non-perturbative effects in the RF systems
are not well understood.

The present paper uses a systematic RG approach. However, some results
can be reproduced more simply with an approximate Migdal-Kadanoff
renormalization group (Appendix B).

The question of the large-distance behavior of the RF and
RA Heisenberg models was discussed in Ref. \cite{AP}
on the basis of an approximate equation of state.
In that paper QLRO was also obtained in the RA case.
However, we believe that this is an accidental
coincidence, since the equation of state \cite{AP} is valid only in
the first order in the strength of the disorder, while higher orders
are crucial for critical properties \cite{G}. In particular, the
approach \cite{AP} incorrectly predicts the absence of QLRO in the
RF XY model and its presence in the exactly solvable RA
spherical model. It also provides incorrect critical exponents in the
Heisenberg case. The reason of the mistakes is the fact that
in the weak external uniform field
the perturbation parameter of Ref. \cite{AP} is large.

The RA Heisenberg model is relevant for the amorphous
magnets \cite{HPZ}. At the same time, for their large-distance
behavior the dipole interaction may be important \cite{B}. Besides, a
weak nonrandom anisotropy is inevitably present due to mechanical
stresses.

In conclusion, we have found that the random anisotropy Heisenberg
model has an infinite correlation length and a power dependence of
the correlation function of the magnetization on the distance at low
temperatures and weak disorder in $4-\epsilon$ dimensions. On the
other hand, the correlation length of the random field $O(N>3)$ model
is always finite.

\acknowledgments

The author thanks E. Domany, G. Falkovich, M.V. Feigelman,
Y. Gefen, S.E. Kor\-shu\-nov, Y.B. Levinson, A.I. Larkin,
V.L. Pokrovsky and A.V. Shytov for useful discussions. This work was
supported by RFBR grant 96-02-18985 and
by grant 96-15-96756 of the
Russian Program of Leading Scientific Schools.

\appendix
\section{INEQUALITY FOR CORRELATION FUNCTIONS}

In this appendix we derive an inequality for the correlation
functions of the disordered systems. We consider the system
with the Hamiltonian

\be
\label{A1}
H=\int dx^D [ H_1(\phi({\bf x})) - h({\bf x})m(\phi({\bf x}))],
\ee
where $\phi$ is the order parameter, $h$ the random field
with short range correlations, $H_1$ may depend on
some other random fields. We prove an inequality for
the Fourier components of the field $m$:

\be
\label{A2}
G_{con}({\bf q})\le
{\rm const}\sqrt{G_{dis}({\bf q})},
\ee
where
$G_{dis}({\bf q})=\overline{\langle{\bf m}({\bf q}){\bf m}(-{\bf q})
\rangle} ,
G_{con}({\bf q})= \overline{\langle{\bf m}({\bf q}){\bf m}(-{\bf
q})\rangle} - \overline{\langle{\bf m}({\bf q})\rangle\langle{\bf
m}(-{\bf q})\rangle}$, the angular brackets denote the thermal
averaging, the bar denotes the disorder averaging.

This inequality can be easily obtained in the case of the Gaussian
distribution $P(h)$ of the field $h$ \cite{SwSo}. Indeed,
in the Gaussian case

\bea
G_{dis}({\bf q})=
\int \left( P(h)
\frac{d}{d h({\bf q})} m_{\bf q}(h)\right) D\{h\}=
& & \nonumber\\
\label{A3}
-\int \left(\frac{d}{dh({\bf q})}P(h)
m_{\bf q}(h)\right) D\{h\}
={\rm const}\int \left(P(h)h(-{\bf q})m_{\bf q}(h)\right)D\{h\},
\eea
where $\int D\{h\}$ denotes the integration over 
the realizations of the random field,
$m_{\bf q}(h)$ $=$
${\int D\{\phi\}\exp(-{H}/{T})m({\bf q})}$ $/$
${\int D\{\phi\}\exp(-{H}/{T})}$.
Applying the Cauchy-Bunyakovsky inequality to Eq. (\ref{A3})
one gets Eq. (\ref{A2}).

However, the assumption about the Gaussian distribution of the random
field is not necessary.
The inequality (\ref{A2}) can also be extended to a
more general situation, corresponding to the
effective replica Hamiltonian (\ref{7}). Indeed,
if one adds to any Hamiltonian
a weak Gaussian random field $\tilde h$ , conjugated with the field $m$,
it suffices for Eq. (\ref{A2}) to become valid. The addition of the
Gaussian random field corresponds to the transformation $R({\bf n}_a{\bf
n}_b)\rightarrow R({\bf n}_a{\bf n}_b)+\Delta{\bf n}_a{\bf n}_b$ in Eq.
(\ref{7}) where $\Delta\sim\tilde h^2$ is a positive constant.  Thus,
Eq. (\ref{A2}) is invalid only, if for the arbitrarily small $\Delta$
the replica Hamiltonians can not contain the two-replica contribution
$\tilde R({\bf n}_a{\bf n}_b) = R({\bf n}_a{\bf n}_b) - \Delta{\bf
n}_a{\bf n}_b$. This corresponds to the border of the region of the
possible Hamiltonians and has zero probability.

For  systems in the critical domain
there is a simple way to understand why the inequality is valid
not only in the Gaussian case but also in the general situation.
This is just a consequence of the universality.

\section{MIGDAL-KADANOFF RENORMALIZATION GROUP}

This appendix contains a simple approximate version of the
renormalization group. The results for the critical exponents of the XY
and Heisenberg models have a very good accuracy. The value of
the magnetization component number $N_c$, at which QLRO
disappears in the RF model, is probably exact. However,
the critical number of the components in the RA model
is underestimated.

\subsection{Random field}

We use the following ansatz for the disorder-induced term
in the Hamiltonian (\ref{1}): $R({\bf n}_a{\bf n}_b)=\alpha{\bf n}_a
{\bf n}_b+\beta$, where $\alpha$ and $\beta$ are constants.
This expression corresponds to the Gaussian RF disorder (\ref{2}).
Below we ignore the generation of the other contributions to
the function $R(z)$. The missed contributions are
related with random anisotropies of different orders.
In terms of the angle variable $\phi$ (\ref{8},\ref{9})

\be
\label{B1}
R(\phi)=\alpha\cos\phi+\beta.
\ee
To ensure consistency we have to truncate the RG equation (\ref{9}).
We substitute the ansatz (\ref{B1}) into Eq. (\ref{9}) and retain
only the terms, proportional to $\cos\phi$ or independent of $\phi$.
This leads to the following RG equation for the constant $\alpha$
(\ref{B1})

\be
\label{B2}
\frac{d\alpha}{d\ln L}=\epsilon\alpha+2\alpha^2(N-3).
\ee
For $N<3$ Eq. (\ref{B2}) has a stable solution
$\alpha=\epsilon/[2(3-N)]$.  The critical exponent (\ref{11}) equals

\be
\label{B3}
\eta=\frac{(N-1)\epsilon}{(3-N)}.
\ee
At $N=2$ this result has less than 
ten percent difference with the systematic
theory \cite{12a}. QLRO disappears at $N=3$. This is the same
critical number which is found in section \ref{sec:IV}.

For $N>3$ a fixed point exists in $4+\epsilon$ dimensions.
It describes the transition between
the ferromagnetic and paramagnetic phases. 
In this fixed point
the critical exponent (\ref{B3}) satisfies the
modified dimensional reduction hypothesis \cite{mdr}. However,
we believe that this is an artifact of the Migdal-Kadanoff
approximation, since the correct value of the critical exponent differs
form Eq. (\ref{B3}).

\subsection{Random anisotropy}

In this case we use the ansatz $R({\bf n}_a{\bf n}_b)=
A({\bf n}_a{\bf n}_b)^2+B$. In terms of the variable
$\phi$ (\ref{8},\ref{9}) $R(\phi)=\alpha\cos2\phi+\beta$.
We again substitute our ansatz into Eq. (\ref{9}) and
retain the terms, proportional to $\cos 2\phi$, and
the terms, independent of $\phi$.
The RG equation for the constant $\alpha$ has the form

\be
\label{B4}
\frac{d\alpha}{d\ln L}=\epsilon\alpha+8(N-6)\alpha^2.
\ee
The fixed point solution of this equation is
$\alpha=\epsilon/[8(6-N)]$.
It describes the QLRO phase at $N<6$.
At $N=3$ the function $R(\phi)=\alpha\cos2\phi+\beta$ is just
$R_{\omega=0}$ of section \ref{sec:V.C}. The
critical exponent of the two-spin correlation function is given by the
following equation

\be
\label{B5}
\eta=\frac{\epsilon(N-1)}{6-N}.
\ee
At $N=2,3$ this value is close to the results of the systematic
approach (Table I).

\section{SPHERICAL MODEL}

In this appendix we consider the spherical RA model
with the functional RG. We show that QLRO is absent in
this model.
In the spherical limit $N=\infty$
only the terms, proportional to $N$, and the term $\epsilon R(z)$
should be retained in the right hand side of Eq. (\ref{Rz}).
After the change of the variable $R(z)=\epsilon r(z)/N$ one obtains

\be
\label{RGr}
0=r(z)
[1+4r'(1)] -
2zr'(1)r'(z) +
(r'(z))^2.
\ee

It is convenient to differentiate Eq. (\ref{RGr}) over $z$.
One gets

\be
\label{difRG}
0=r'(z)[1+2r'(1)]+2r''(z)[r'(z)-zr'(1)].
\ee
Analytical functions $r(z)$ can satisfy Eq. (\ref{difRG}) at $z=1$
only if $r'(1)=0$ or $r'(1)=-1/2$. In both cases Eq. (\ref{difRG})
can be easily solved. There are three analytical non-zero solutions:
$r(z)=-z/2+1/4; r(z)=-(1-z)^2/4; r(z)=-z^2/4$. The last
solution only has the necessary symmetry.

The non-analytical solutions are absent.
Indeed, Eq. (\ref{difRG}) can be integrated with the substitution
$r'(z)=zt(z)$. The general integral has the form

\be
\label{genint}
\frac{(t(z))^{2t(1)}}{(2t(z)+1)^{2t(1)+1}}=Cz.
\ee
Besides, there are special solutions.  They all satisfy
the relation $t(z)=t(1)$.
Hence, the special
solutions are analytical. Thus, the function $t(z)$ can
be non-analytical at $z=1$ only under the condition that $z=1$ is a
peculiar point of Eq. (\ref{genint}). This means that $t(1)=0$ or
$t(1)=-1/2$.  However, it is easy to verify that
in both cases the solution is one of the found above.

We see that the only fixed point of the spherical RA model
is $R(z)=-\epsilon z^2/(4N)$. With Eq. (\ref{11}) one finds
the critical exponent $\eta=-\epsilon/2$. Since $\eta>0$
the solution found is applicable at $D>4$. At $D<4$
the fixed points are absent. Thus, QLRO is absent too.

%\end{thebibliography}

\begin{references}
%\begin{thebibliography}{99}

\bibitem{IM}
Y. Imry and S.K. Ma, Phys. Rev. Lett. {\bf 35}, 1399 (1975).

\bibitem{HPZ}
R. Harris, M. Plischke, and M.J. Zuckermann,  Phys. Rev. Lett.
{\bf 31}, 160 (1973).

\bibitem{OS}
D.J. Sellmyer and M.J. O'Shea,  in {\it Recent
Progress in Random Magnetism}, ed. D. Ryan (World Scientific,
Singapore, 1992) p. 71.

\bibitem{LQ}
N.A. Clark, T. Bellini, R.M. Malzbender, B.N. Thomas,
A.G. Rappaport, C.D. Muzny, D.W. Shaefer, and L. Hrubesh, 
Phys. Rev. Lett. {\bf 71}, 3505 (1993).
T.Bellini, N.A. Clark, and D.W. Schaefer, Phys. Rev. Lett.
{\bf 74}, 2740 (1995).
H. Haga and C.W. Garland, Liq. Cryst. {\bf 22}, 275 (1997).

\bibitem{NE}
S.V. Fridrikh and E.M. Terentjev, Phys. Rev. Lett. {\bf 79}, 4661
(1997). T. Emig, Phys. Rev. Lett. {\bf 82}, C3380 (1999).

\bibitem{He3}
J.V. Porto III and J.M. Parpia, Phys. Rev. Lett. {\bf 74}, 4667
(1995). 
K. Matsumoto, J.V. Porto, L. Pollak, E.N. Smith, T.L. Ho, 
and J.M. Parpia, 
Phys. Rev. Lett. {\bf 79}, 253
(1997).

\bibitem{HTSC}
G. Blatter, M.V. Feigel'man, V.B. Geshkenbein, A.I. Larkin,
V.M. Vinokur, 
Rev. Mod. Phys. {\bf 66}, 1125 (1994).

\bibitem{L}
A.I. Larkin, Zh. Eksp. Teor. Fiz. {\bf 58}, 1466 (1970)
[Sov. Phys. JETP {\bf 31}, 784 (1970)].

\bibitem{P}
R.A. Pelcovits, E. Pytte, and J. Rudnik, Phys. Rev. Lett. {\bf 40},
476 (1978).

\bibitem{AW}
M. Aizeman and J. Wehr, Phys. Rev. Lett. {\bf 62}, 2503 (1989);
Commun. Math. Phys. {\bf 150}, 489 (1990).

\bibitem{Xray}
U. Yaron, P.L. Gammel, D.A. Huse, R.N. Kleiman, C.S. Oglesby,
E. Bucher, B. Batlogg, D.J. Bishop, 
K. Mortensen, K. Clausen, C.A. Bolle,
and F. De La Cruz, 
Phys. Rev. Lett. {\bf 73}, 2748 (1994).

\bibitem{RFXY}
S.E. Korshunov, Phys. Rev. B {\bf 48}, 3969  (1993).

\bibitem{12a}
T. Giamarchi and P. Le Doussal, Phys. Rev. Lett. {\bf 72}, 1530
(1994); Phys. Rev. B {\bf 52}, 1242 (1995).

\bibitem{numXY}
M.J.P. Gingras and D.A. Huse, Phys. Rev. B {\bf 53}, 15193 (1996).

\bibitem{FRG}
D.S. Fisher, Phys. Rev. Lett. {\bf 56}, 1964 (1986).
L. Balents and D.S. Fisher, Phys. Rev. B {\bf 48}, 5949 (1993).

\bibitem{MP}
M. Mezard and G. Parisi, J. Phys. A {\bf 23},
L1229 (1990); J. Phys. I France {\bf 1}, 809 (1991).

\bibitem{F}
D.E. Feldman, Pis'ma ZhETF {\bf 65}, 108 (1997) [JETP Lett. {\bf 65}, 114
(1997)]; Phys. Rev. B {\bf 56}, 3167 (1997).

\bibitem{EN}
T. Emig and T. Nattermann, Phys. Rev. Lett. {\bf 81}, 1469 (1998);
cond-mat/9810367. A. Hazareesing and J.-P. Bouchaud,  cond-mat/9810097.

\bibitem{RT}
L. Radzihovsky and J. Toner, cond-mat/9811105.

\bibitem{AP}
A. Aharony and E. Pytte, Phys. Rev. Lett. {\bf 45}, 1583 (1980).

\bibitem{B}
B. Barbara, M. Coauch, and B. Dieny, Europhys. Lett. {\bf 3}, 1129
(1987).

\bibitem{num}
R. Fisch, Phys. Rev. B {\bf 57}, 269 (1998); {\it ibid} {\bf 58},
5684 (1998).
J. Chakrabaty, Phys. Rev. Lett. {\bf 81}, 385 (1998).

\bibitem{spher}
P. Lacour-Gayet and G. Toulouse, J. Phys. (Paris) {\bf 35}, 425 (1974).
S.L. Ginzburg, Zh. Eksp. Teor. Fiz. {\bf 80}, 244 (1981).
A. Khurana, A. Jagannathan, and J.M. Kosterlitz, Nucl. Phys. B
{\bf 240}, 1 (1984). M.V. Feigelman and M.V. Tsodyks. Zh. Eksp.
Teor. Fiz. {\bf 91}, 955 (1986) [Sov. Phys. JETP {\bf 64}, 562 (1986)].

\bibitem{G}
Y.Y. Goldshmidt, Nucl. Phys. B {\bf 225},
123 (1983).

\bibitem{SwSo}
M. Schwartz and A. Soffer, Phys. Rev. Lett. {\bf 55}, 2499 (1985).

\bibitem{Pol}
A.M. Polyakov, Phys. Lett. {\bf 59B}, 79 (1975);
{\it Gauge Fields and Strings} (Harwood Academic Publishers, Chur,
1987).

\bibitem{prim}
The results about QLRO, obtained for the RF XY model in Ref.
\cite{RFXY,12a}, can be easily extended to the RA XY model, since in
terms of the angles $\phi({\bf x})$ between the spins ${\bf n}({\bf x})$
and some fixed direction the Hamiltonians of these models are almost
identical.

\bibitem{DF}
D.S. Fisher, Phys. Rev. B {\bf 31}, 7233 (1985).

\bibitem{ZJ} J. Zinn-Justin, {\it Quantum Field Theory and Critical
Phenomena} (Oxford University Press, Oxford, 1993).

\bibitem{2DGSsim}
C. Zeng, P.L. Leath, and D.S. Fisher, cond-mat/9807281.

\bibitem{vortex}
 T. Garel, G. Iori, and H. Orland, Phys. Rev. B {\bf 53},
 R2941 (1996). D. Carpentier, P. Le Doussal, and
 T. Giamarchi, Europhys. Lett. {\bf 35}, 379 (1996).
 J. Kierfeld, T. Nattermann, and T. Hwa,
 Phys. Rev. B {\bf 55}, 626 (1997).

\bibitem{mdr}
M. Schwartz and A. Soffer, Phys. Rev. B {\bf 33}, 2059 (1986).

\end{references}
\end{document}